\documentclass{ifacconf}

\usepackage{graphicx,color}      
\usepackage{natbib}        
\usepackage{amssymb,amsmath}
\usepackage{subfigure}

\usepackage{etoolbox}
\AtEndEnvironment{thm}{\hfill $\triangleleft$}
\AtEndEnvironment{prob}{\hfill $\triangleleft$}
\AtEndEnvironment{pf}{\hfill $\diamond$}
\AtEndEnvironment{lemma}{\hfill $\triangleleft$}
\AtEndEnvironment{proposition}{\hfill $\triangleleft$}
\AtEndEnvironment{remark}{\hfill $\triangleleft$}

\newcommand{\argmax}{\mathop{\rm arg~max}\limits}

\usepackage{multicol}

\newcommand{\R}{{\mathbb R}}

\newcommand{\Rnn}{\R ^{n\times n}}

\newcommand{\calC}{{\mathcal C}}

\newcommand{\calE}{{\mathcal E}}
\newcommand{\calF}{{\mathcal F}}

\newcommand{\calS}{{\mathcal S}}

\newcommand{\calV}{{\mathcal V}}
\newcommand{\calW}{{\mathcal W}}
\newcommand{\calX}{{\mathcal X}}

\newcommand{\bbE}{{\mathbb E}}

\newcommand{\KL}[2]{D_{\mathrm{KL}} \left( {#1} \middle\| {#2} \right)}

\def\part#1#2{\frac{\partial #1}{\partial #2}}
\def\vec#1{\mbox{\boldmath $#1$}}

\begin{document}
\begin{frontmatter}

\title{Resilience Evaluation of Entropy Regularized Logistic Networks with Probabilistic Cost}

\author[TC]{Koshi Oishi}
\author[KK]{Yota Hashizume}
\author[TO]{Tomohiko Jimbo}
\author[TO]{Hirotaka Kaji}
\author[KK]{Kenji Kashima}

\address[TC]{K. Oishi is affiliated with the Toyota Central R\&D LABS., inc., Aichi 480-1192, Japan. \tt{e1616@mosk.tytlabs.co.jp}}
\address[TO]{T. Jimbo and H. Kaji are affiliated with the Frontier Research Center, Toyota Motor Corporation, Aichi, 471-8571, Japan.}
\address[KK]{Y. Hashizume and K. Kashima are affiliated with the Graduate School of Informatics, Kyoto University, Kyoto 606-8501, Japan.}

\begin{abstract}                
The demand for resilient logistics networks has increased because of recent disasters.
When we consider optimization problems, entropy regularization is a powerful tool for the diversification of a solution.
In this study, we proposed a method for designing a resilient logistics network based on entropy regularization.
Moreover, we proposed a method for analytical resilience criteria to reduce the ambiguity of resilience.
First, we modeled the logistics network, including factories, distribution bases, 
and sales outlets in an efficient framework using entropy regularization.
Next, we formulated a resilience criterion based on probabilistic cost and Kullback--Leibler divergence.
Finally, our method was performed using a simple logistics network, and the resilience of the three logistics plans designed by entropy regularization was demonstrated.
\end{abstract}


\end{frontmatter}
%
%
%
\section{Introduction}

Supply chains (SCs) are important networks that support modern society, and companies of all sizes continuously seek to improve their economy and efficiency.
However, SCs are vulnerable to disruptions in logistics networks and
fluctuations in demand owing to the recent COVID-19 pandemic and natural disasters.
Therefore, the transformation to SC resilience (SCR) is necessary; see, e.g., \cite{Forum2022, Trade2021,davis2021towards}.

SCR has been studied since 2000, and many studies have been reported to date; see, e.g., \cite{han2020}.
For example, \cite{hatefi2014} proposed a design method for a robust logistics network using a mixed-integer linear programming model with augmented unexpected destruction constraints.
\cite{Sergey2012,Sergey2012_Raz} applied a control theory framework to logistics networks to derive distributed control rules for disaster-resistant logistics.
However, these studies differ in their focus on resilience.
\cite{han2020} surveyed research on SCR and classified performance metrics of SCR into 11 categories, 
such as ``Performance in maintaining customer satisfaction'' and ``Efficiency in recovering to normality.''
In this study, we focused on the damage caused by disruptions, which refers to resilience against destruction caused by external factors.

In this study,
we focused on the resilience of the logistics network, which is an important element of SCR.
In particular, we utilized entropy-regularized optimization for designing resilient networks because this method can provide flexibility; see \cite{mnih2016,haarnoja2018soft,eysenbach2021}.
In fact, our attempt is an important first step toward viewing the logistic network as a control of distributions; see \cite{kk2022}. 
The contributions of this study are as follows:

\begin{itemize}
     \item Modeling the optimization problem with entropy regularization for 3-layer logistics networks.
     \item Converting the aforementioned problem to Schr\"{o}dinger bridge for efficient calculation.
     \item Proposing an analytical evaluation method for logistics networks.
     \item Demonstrating resilience via numerical simulations using the proposed method.
\end{itemize}
   
The remainder of this paper is organized as follows.
In Section \ref{sec:model}, we model the logistics network as an optimization problem with entropy regularization.
In Section \ref{sec:ev}, we propose a resilience evaluation method for the model described in Section \ref{sec:model}.
Simulations of resilience evaluation using a simple logistics network as an example are presented in Section \ref{sec:sim}.
Section \ref{sec:concl} concludes the paper.

\section{Modeling and optimization}
\label{sec:model}

A basic logistics network consists of factories, distribution bases (DBs), and sales outlets (see, for example, \cite{nozick1998,darmawan2021}).
In this section, the network design is formulated as entropy-regularized optimization.

\subsection{Logistic network design with entropic optimization}
\label{sec:design}

\begin{figure}[t]
     \centering
     \includegraphics[width=65mm]{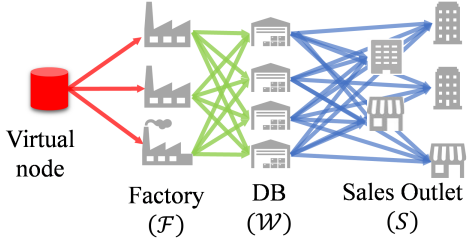}
     \caption{Bidirectional graph representation of a logistic network.}
     \label{fig:g}
\end{figure}

We consider optimizing production quantities and transportation paths for a logistics network with factories, DBs, and sales outlets.
Let $\mathcal{F}, \ \mathcal{W} \ \text{and} \ \mathcal{S}$ denote the (finite) set of labels of factories, DBs, and sales outlets. For notational simplicity,
\begin{align}
     \mathcal{V} := \{\rm i\}\cup\mathcal{F}\cup\mathcal{W}\cup\mathcal{S},
\end{align}
where $\rm i$ denotes a virtual node, representing the production quantities. As depicted in Fig.~\ref{fig:g},
\begin{align}
   \mathcal{E} := (\{\rm i\}\times {\mathcal {F}})\cup (\mathcal{F} \times \mathcal{W})\cup(\mathcal{W}\times \mathcal{S}).
\end{align}
For each edge $e\in \calE$, its cost $\vec{A}_e\in\R$ is given. 
For example, $\vec{A}_{({\rm i},f)},\ \vec{A}_{(f,w)}, \text{and} \ \vec{A}_{(w,s)}$ respectively represent the production cost at factory $f\in \calF$, transport cost from factory $f$ to DB $w\in \calW$, including storage cost, and transport cost from DB $w$ to sales outlet $s$.
The set of paths is denoted as $\calX:=\calF\times\calW\times \calS$.
The cost of path $x=(f,w,s)$ is given by
\begin{align}\label{eq:Cx}
     \vec{C}_x := \vec{A}_{({\rm i},f)}+ \vec{A}_{(f,w)} + \vec{A}_{(w,s)}.
\end{align}
The goal is to determine a logistic plan $P_\cdot:\calX\to \R$ such that
\begin{align}
   \sum_{x\in\calX} P_x =1,\ P_x\ge 0, \label{eq:const1}\\
   \sum_{x\in \calX_s} P_x = \zeta_s,\ \forall s\in\calS, \label{eq:const2}
\end{align}
where $\calX_s \subset \calX \; \text{for any} \; s\in \calS$ is the set of paths to sales outlet $s$ (i.e., $\calX_s:=\{(f,w,s):f\in\calF,\ w\in\calW\}$) and $\zeta_s$ is the distribution of the demand quantity satisfying
\begin{align}
  \sum_{s\in\calS} \zeta_s =1,\ \zeta_s\ge 0. \label{eq:zeta}
\end{align}
\begin{prob}%
  \label{prob:opt}
  Using the aforementioned notation, and given $\zeta_s\ (s\in\calS)$ and $\alpha>0$, we find $P$ that minimizes
  \begin{align}
    \sum_{x\in\calX} \vec{C}_{x} P_{x} - \alpha \mathcal{H}(P),
    \label{eq:opt}
  \end{align}
  satisfying \eqref{eq:const1} and \eqref{eq:const2}, where $\mathcal{H}(P)$ denotes the entropy of plan $P$:
  \begin{equation*}
      \mathcal{H}(P) := -\sum_{x\in \calX} P_{x} \log{P_{x}}.
      \label{eq:entropy}
  \end{equation*}
\end{prob}
 
From (\ref{eq:opt}), Problem \ref{prob:opt} becomes a general cost-minimization problem when $\alpha$ is small. 
However, Problem \ref{prob:opt} becomes an optimization problem that increases entropy $\mathcal{H}(P)$ when $\alpha$ is large.
We consider that a large $\mathcal{H}(P)$ implies that $P$ approaches a uniform distribution.
Therefore, $\alpha$ can be used to adjust the trade-off between the economics and flexibility of a logistics plan.

\subsection{Efficient calculation via Schr\"{o}dinger bridge}
\label{sec:sh}

As the number of nodes $|\mathcal{V}|$ increases, using a nonlinear solver to solve Problem (\ref{prob:opt}) becomes unrealistic.
We circumvent this computational issue by converting Problem \ref{prob:opt} to the so-called Schr\"{o}dinger bridge problem.
\begin{thm}
     Problem \ref{prob:opt} is equivalent to the Schr\"{o}dinger bridge of finding $P$ that minimizes
     \begin{eqnarray}
          \begin{split}
            & \KL{P}{\mathfrak{M}_{\rm RB}}\\
          \end{split}
          \label{eq:sh}
        \end{eqnarray}
        satisfying \eqref{eq:const1} and \eqref{eq:const2},
        where $\mathfrak{M}_{\rm RB}$ is the distribution based on the Ruelle-Bowens random walk (see \cite{delvenne2011}) and $D_{\rm KL}$ is the Kullback--Leibler (KL) divergence.
\label{the:sh}
\end{thm}
\begin{pf}
     See Appendix \ref{app:sh}.
\end{pf}

\begin{figure}[t]
     \centering
     \includegraphics[width=80mm]{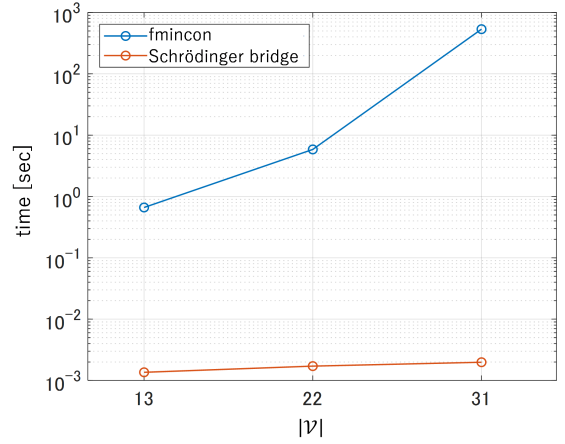}
     \caption{Computation time for each $|\mathcal{V}|$ case. The vertical axis is a logarithmic scale.}
     \label{fig:time}
\end{figure}

An efficient iterative algorithm is available for solving the Schr\"{o}dinger bridge problem; see \cite{chen2016robust,chen2017efficient}.
Fig.~\ref{fig:time} shows the average computation time of 10 iterations for Problem \ref{prob:opt}
with the `{\tt fmincon}' function of MATLAB\textsuperscript{\tiny\textregistered} R2021b and with Schr\"{o}dinger bridge.
In Fig.~\ref{fig:time}, $\mathcal{F}$, $\mathcal{W}$, and $\mathcal{S}$ increase by three.
In other words, the total number of nodes $\mathcal{V}$ increases by nine.
The positions of the nodes that affect edge costs are randomized.
The `{\tt fmincon}' setting is the default.
A laptop computer with a 16-core Intel\textsuperscript{\tiny\textregistered} Xeon\textsuperscript{\tiny\textregistered} Gold 6242 (2.80 GHz) processor and 32 GB RAM was used for the calculations.

\section{Evaluation of logistics network resilience}
\label{sec:ev}

Many studies have been conducted on resilience in the field of logistics, although the evaluation differs for each study; see \cite{han2020}.
In this study, resilience was evaluated based on its performance against the damage caused by disruptions.
Specifically,
we quantified the scale of disruption caused by KL divergence,
and considered the worst total cost of the logistics network at this scale of disruption.
Additionally,
it is important to evaluate each edge of the graph, which corresponds to the production quantity and paths between bases in logistics networks.
Therefore, we focus on cost fluctuations per edge.

To this end, we introduce probabilistic cost $A_\cdot$, which is a random variable.
Then, the set of probabilistic costs $\tilde A$ that fluctuates from nominal $A$ is defined as
\begin{equation}
     \calC (A, \epsilon) =
     \left\{ \tilde A
     : \sum_{e\in \calE} D_{\rm KL} ( {\tilde A}_e ||  A_e ) \leq \epsilon
     \right\},
\label{eq:set}
\end{equation}
for $\epsilon >0$, where $D_{\rm KL} ( {\tilde A}_e ||  A_e ) $ denotes the KL divergence between the two real random variables ${\tilde A}_e,\ A_e$.


Next, we investigated the total cost for this fluctuating cost class.
We propose the worst total cost in $ \mathcal{C}$ as the evaluation metric.
The expected value of the total cost is as follows:
\begin{equation}
     L(A,P) := \bbE_{\tilde A}\left[\sum_{x\in\calX} C_x(\tilde A)P_x \right] ,
     \label{eq:c_total}
\end{equation}
with random variable $C_x$ defined by
\begin{align}\label{eq:Cx2}
    C_x(\tilde A) := \tilde A_{({\rm i},f)}+ \tilde A_{(f,w)} + \tilde A_{(w,s)}.
\end{align}
We note that the expectation is considered with respect to the probabilistic uncertainty in $\tilde A$.
From (\ref{eq:set}) and (\ref{eq:c_total}),
the problem of finding the maximum total cost with the scale of disruption $\epsilon$ is formulated as the following optimization:
\begin{prob}     \label{prob:cost}
     Under the same notation as in Section \ref{sec:model}, we suppose that $A$ and $\epsilon$ are given.
     Then, we find
     \begin{align}          \label{eq:allCost}
        L^*(A,\epsilon,P):=\max_{\tilde{A}\in {\mathcal{C}(A,\epsilon)}} L(\tilde{A},P).
     \end{align}
\end{prob}

Obtaining the expected value of (\ref{eq:Cx2}) requires as many integral computations as the number of edges.
It is preferable to solve Problem \ref{prob:cost} explicitly because the number of edges in the general logistics network tends to increase.
Thus, we show that Problem \ref{prob:cost} has an explicit solution.
\begin{thm}
     Suppose that $\{A_e\}_{e\in\calE}$ are independent normal distributions so that
        \begin{equation}
             A_e \sim \mathcal{N} ( \vec{A}_e, \vec{\sigma}_e^2), \\
             \label{eq:normal}
        \end{equation}
        where $\vec{A}_e \in \mathbb{R}$ is the mean and $\vec{\sigma} >0$ is the variance.
        Define the edge occupation probability $\vec{\phi}^P$ for logistic plan $P$ by
        \begin{eqnarray}
             \vec{\phi}^P_e := \frac{1}{3} \sum_{ x \in \mathcal{X}_e} P_x
             \label{eq:varphi}
        \end{eqnarray}
        where $\calX_e \subset \calX \; \text{for any} \; e\in \calE$ is the set of paths containing edge $e$.
        Then, the worst-case cost in Problem \ref{prob:cost} is expressed by
        \begin{equation}
             L^*(A,\epsilon,P) = \sum_{e \in \mathcal{E}} \vec{\phi}^P_e \vec{A}_e +
             \left(2\epsilon \sum_{e \in \mathcal{E}} (\vec{\phi}_{e}^{P})^2 \vec{\sigma}_e^2 \right)^{\frac{1}{2}} .\\
             \label{eq:costSol2}
        \end{equation}
        Moreover,
        $A_e^*$, which provides $L^*(A,\epsilon,P)$, follows a normal distribution.
        Then, the mean $\vec{A}^*_e$ of $A_e^*$ can be expressed as follows:
        \begin{equation}
            \vec{A}^*_e = \vec{A}_e + \vec{\sigma}_e^2 \vec{\phi}_{e}^{P} \sqrt{\frac{2\epsilon}{\sum_{e' \in \mathcal{E}} (\vec{\phi}_{e'}^{P})^2 \vec{\sigma}_{e'}^2}} .
        \label{eq:mean}
        \end{equation}
        \label{the:cost}
\end{thm}

\begin{figure}[t]
     \centering
     \includegraphics[width=80mm]{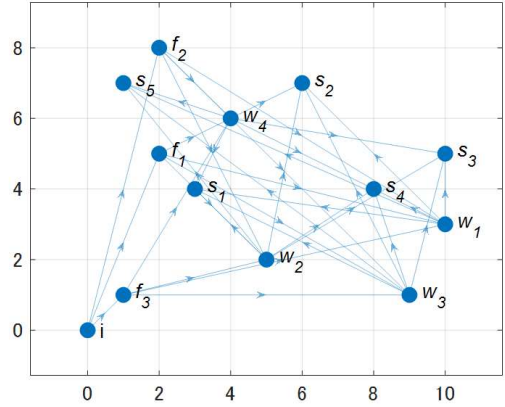}
     \caption{Target logistics network}
     \label{fig:net}
\end{figure}

\begin{figure*}[t]
     \begin{center}
       \subfigure[$\alpha=0.3$]{
               \includegraphics[width=.315\linewidth]{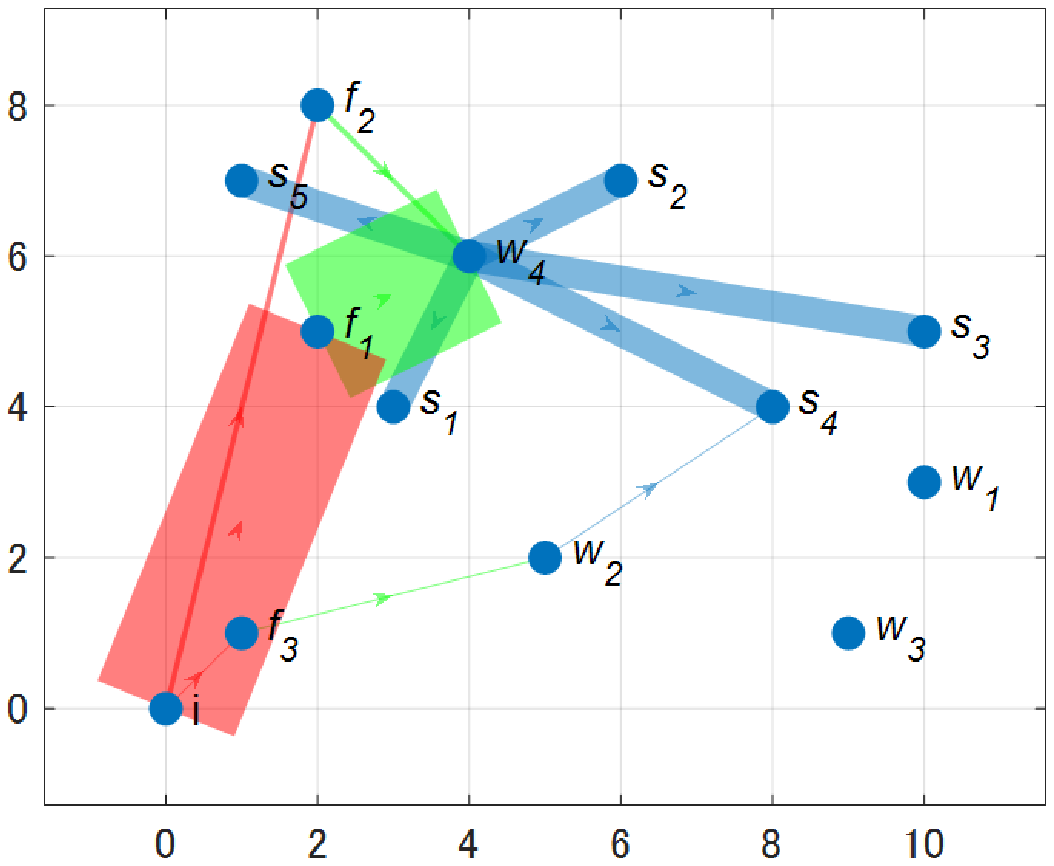}
         \label{fig:opt1}
       }
       \subfigure[$\alpha=0.9$]{
               \includegraphics[width=.315\linewidth]{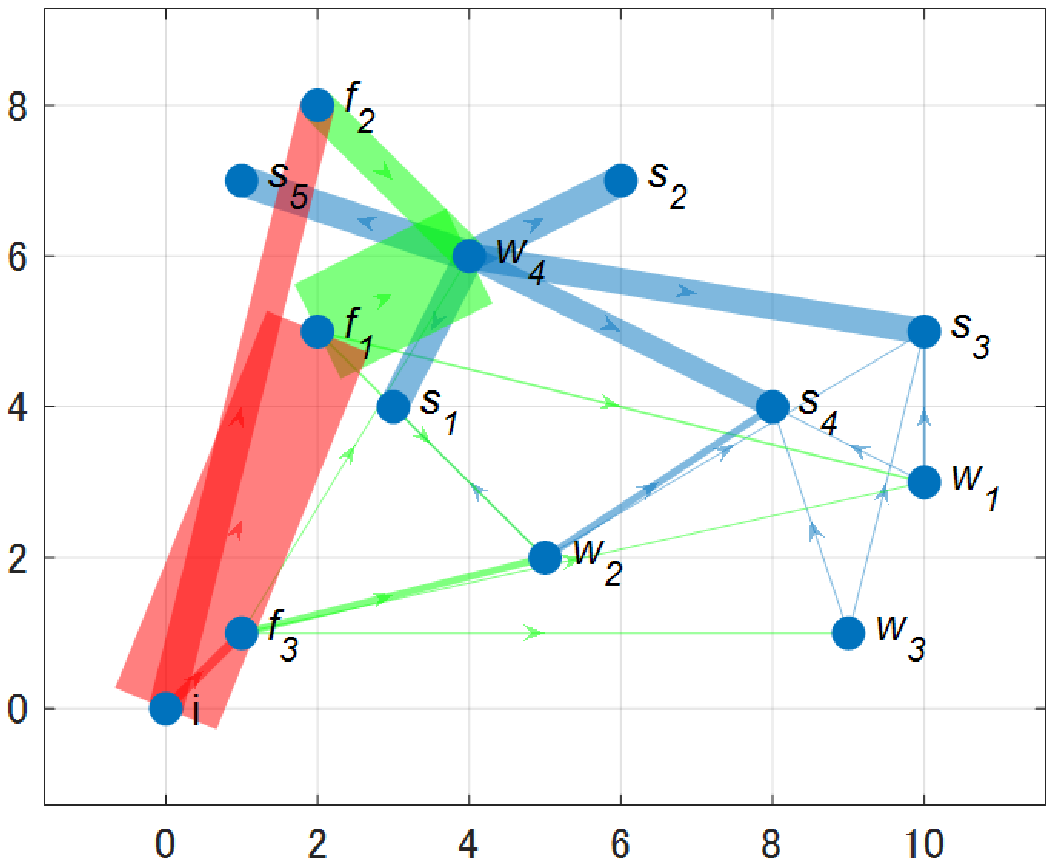}
         \label{fig:opt2}
       }
       \subfigure[$\alpha=7.0$]{
               \includegraphics[width=.315\linewidth]{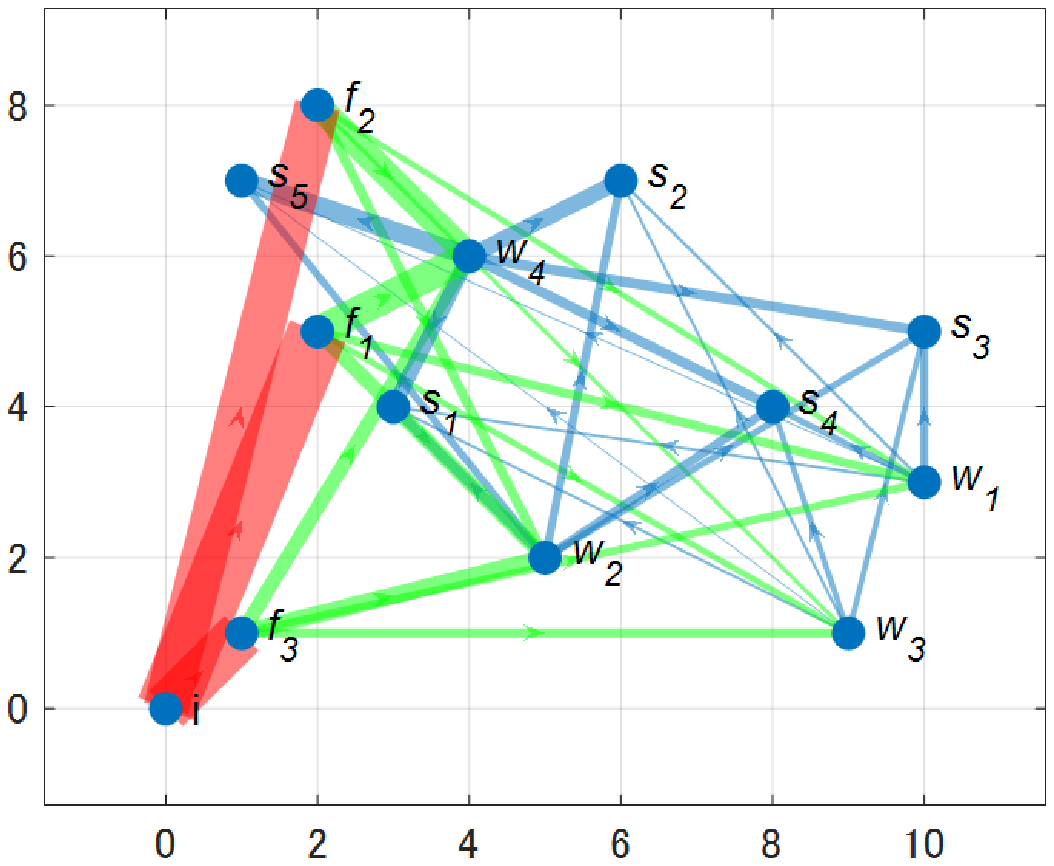}
         \label{fig:opt3}
       }
     \end{center}

     \caption{Logistics plan at each $\alpha$.}
     \label{fig:opt}
\end{figure*}

\begin{pf}
     The probability density function of $A_e$ is defined as $p(A_e)$, and (\ref{eq:c_total}) is rearranged as follows:
     \begin{align}
          L^*(A,P) = \sum_{e \in \mathcal{E}} \phi^P_e \int p(A_e) A_e d A_e.
          \label{eq:A_e}
     \end{align}
     According to the Lagrange multiplier method, the min-max theorem, and (\ref{eq:A_e}), 
     the maximum value $L^*(A,\epsilon,P)$ of (\ref{eq:allCost}) is rearranged as follows:
     \begin{eqnarray}
          \begin{split}
          & L^*(A,\epsilon,P)  = \\
          & \; \min_{\tau \geq 0} \left\{ \tau \epsilon + \tau \sum_{e \in \mathcal{E}}
          \log \int p(A_e) \exp{\left(\frac{\vec{\phi}^P_e}{\tau} A_e \right) } \, d A_e \right\}.
          \end{split}
          \label{eq:costSol}
     \end{eqnarray}
     From (\ref{eq:normal}), the antilogarithm of (\ref{eq:costSol}) is expressed as follows:
     \begin{eqnarray*}
          \begin{split}
               & \int p(A_e) \exp{\left(\frac{\vec{\phi}_{e}^{P}}{\tau} A_e\right)} \, d A_e \\
               & = \int \frac{1}{\sqrt{2 \pi \sigma^2}} \exp{\left(- \frac{(A_e - \vec{A}_e)^2}{2 \sigma^2} \right)} \exp{\left(\frac{\vec{\phi}_{e}^{P}}{\tau} A_e \right)} d A_e \\
               & = \exp{\left(\vec{A}_e \frac{\vec{\phi}_{e}^{P}}{\tau} + \frac{(\vec{\phi}_{e}^{P})^2 \sigma^2}{2\tau^2}\right)} \\
               & \int \frac{1}{\sqrt{2 \pi \sigma^2}} \exp{\left[- \frac{1}{2 \sigma^2} \left\{A_e - \left(\vec{A}_e + \frac{\vec{\phi}_{e}^{P} \sigma^2}{\tau} \right) \right\} \right] }^2 d A_e \\
               & = \exp{\left( \vec{A}_e \frac{\vec{\phi}_{e}^{P}}{\tau} + \frac{(\vec{\phi}_{e}^{P})^2 \sigma^2}{2 \tau^2} \right)}.
          \end{split}
          \label{eq:log}
     \end{eqnarray*}
     Therefore, (\ref{eq:costSol}) is given as follows.
     \begin{eqnarray*}
          \begin{split}
          & \min_{\tau \geq 0} \left\{ \tau \epsilon + \tau \sum_{e \in \mathcal{E}}
          \log \int p(A_e) \exp{\left(\frac{\vec{\phi}_{e}^{P}}{\tau} A_e \right)} \, d A_e \right\} \\
          & = \min_{\tau \ge 0} \left\{ \tau \epsilon + \tau \sum_{e \in \mathcal{E}} \left( \vec{A}_e \frac{\vec{\phi}_{e}^{P}}{\tau} + \frac{(\vec{\phi}_{e}^{P})^2 \sigma^2}{2\tau^2}\right) \right\} \\
          & = \sum_{e \in \mathcal{E}} \vec{\phi}_{e}^{P} \vec{A}_e + \min_{\tau \ge 0} \left\{ \tau \epsilon + \frac{1}{\tau} \sum_{e \in \mathcal{E}} \frac{(\vec{\phi}_{e}^{P})^2 \sigma^2}{2} \right\} \\
          & = \sum_{e \in \mathcal{E}} \vec{\phi}_{e}^{P} \vec{A}_e + \left(2 \epsilon \sum_{e \in \mathcal{E}} (\vec{\phi}_{e}^{P})^2 \sigma^2 \right)^{\frac{1}{2}}.
          \end{split}
          \label{eq:costSol3}
     \end{eqnarray*}  

     \label{pr:cost}
\end{pf}

We consider the evaluation of resilience using $L^*(A,\epsilon,P)$ in (\ref{eq:costSol2}).
Resilience and robustness are similar measures; see, e.g., \cite{prorok2021}.
Robustness is often used as a measure of flexibility against single-scale disturbances.
By contrast, resilience is used as a measure of flexibility against diverse scale disturbances.
Our method can vary the upper bound of the disruption scale with uncertainty.
Therefore, we propose this as a criterion for resilience.

\begin{figure}[t]
     \centering
     \includegraphics[width=70mm]{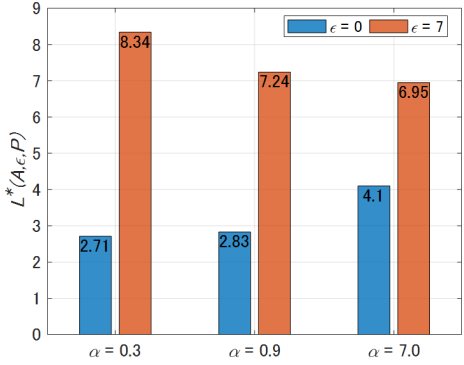}
     \caption{Worst total costs $L^*(A,\epsilon,P)$ of each $\alpha$ in $\epsilon=0$ and $\epsilon=7$.}
     \label{fig:total_cost}
\end{figure}

\section{Numerical simulation}
\label{sec:sim}

To confirm the effectiveness of the proposed evaluation method,
numerical simulations were performed for a simple logistics network involving factories, DBs, and sales outlets.

\subsection{Condition}
\label{sec:cond}

The target logistics network with $|\mathcal{F}| = 3$, $|\mathcal{W}| = 4$, and $|\mathcal{S}| = 5$ is shown in Fig.~\ref{fig:net}.
Here, $f_i \; (i=1, \dots,3)$, $w_i \; (i=1,\dots,4)$, and $s_i \; (i=1,\dots,5)$ denote elements of $\mathcal{F}$, $\mathcal{W}$, and $\mathcal{S}$, respectively.
Furthermore, $\rm i$ is the virtual node described in Section \ref{sec:design} and each axis in Fig.~\ref{fig:net} illustrates this position.
We assumed that $\vec{A}_{({\rm i}, f)}$ for all $f$, which denotes the production costs, are the same.
Therefore, the distances between $\rm i$ and $f$ were irrelevant in Fig.~\ref{fig:net}.
Moreover, the distribution of demand quantity $\zeta_s$ is given by
\begin{align*}
     \zeta_s = \frac{1}{5},  \forall s \in \mathcal{S} .
\end{align*}
\begin{figure}[t]
     \centering
     \includegraphics[width=70mm]{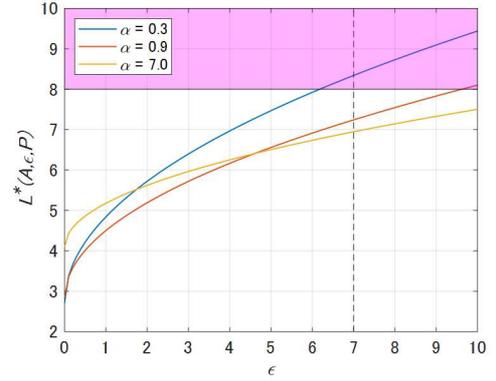}
     \caption{Worst total costs $L^*(A,\epsilon,P)$ of each $\alpha$ in each $\epsilon$; filled areas: $L^*(A,\epsilon,P) \ge 8$.}
     \label{fig:total_cost_ep}
\end{figure}

\subsection{Design of logistics plan}
\label{sec:logi}

The three logistics plans with different parameters $\alpha$ are shown in Fig.~\ref{fig:opt}.
The width of the edge denotes the occupation probability $\vec{\phi}_{e}^{P}$.
The red, green, and blue lines denote the edges of $({\rm i},f)$, $(f,w)$, and $(w,s)$, respectively. 
We note that edges with quantities of 1\% or less are not shown.
As shown in Fig.~\ref{fig:opt1}, for the smallest $\alpha (=0.3)$, the low-cost path from $\rm i$ to $s$ is selected.
Focusing on the location of each node,
the path from $f_1$ to $s$ via $w_4$ has the lowest cost.
Therefore, as a priority cost,
the plan is designed to produce a product with $f_1$ and transport it to $s$ via $w_4$.
The other plan with $\alpha=0.9$ not only uses lowest-cost paths but also uses a high-cost path for transporting a small quantity of products, as shown in Fig.~\ref{fig:opt2}.
For the largest $\alpha(=7.0)$, the production quantity is nearly even, as shown in Fig.~\ref{fig:opt3}.
Moreover, Fig.~\ref{fig:opt3} is planned to contain more diverse transportation paths than those in Fig.~\ref{fig:opt2}.
Therefore,
the trade-off between cost and flexibility can be considered using the method described in Section \ref{sec:design}.

\subsection{Evaluation of resilience}
\label{sec:compa}

The three logistics plans in Section \ref{sec:logi} were evaluated in terms of resilience.
Fig.~\ref{fig:total_cost} shows the $L^*(A,\epsilon,P)$ of (\ref{eq:costSol2}) with disruptions $\epsilon(=0,7)$.
The logistics plan with $\alpha=0.3$ has the lowest cost in the case of $\epsilon = 0$.
However, the logistics plan designed with $\alpha=0.3$ has the highest cost when $\epsilon = 7$.
Assuming that the $L^*(A,\epsilon,P)$ greater than $8$ renders business impossible,
we can observe that the logistics plans designed with $\alpha=0.9$ and $\alpha=7.0$ are robust for the upper bound.

In this study, the robustness of the variable upper bounds is defined as resilience. 
The evaluation was performed when $\epsilon$ was increased, as shown in Fig.~\ref{fig:total_cost_ep}.
Our method can be evaluated analytically even if $\epsilon$ is increased significantly.
Fig.~\ref{fig:total_cost_ep} shows $L^*(A,\epsilon,P)$ for each logistics plan of $\alpha$.
As in the evaluation of robustness, we assume that the $L^*(A,\epsilon,P)$ greater than $8$ renders the business impossible.
We can observe that only $\alpha=7.0$ can allow the business to continue, even if $\epsilon=10$.
Thus, this result suggests that the logistics plan with $\alpha=7.0$ is more resilient than that with $\alpha= 0.9$.

\subsection{Resilience for single-edge disruption}
\label{sec:edge}

The disruption of a single edge is an easily imaginable event in logistics networks. To investigate this incident, we introduce
single-edge disruption at edge $e$ using
\begin{equation}
     \calC_e (A, \epsilon) :=
     \left\{ \tilde A \in \calC (A, \epsilon)
     : \tilde A_{e'} = A_{e'},\ \forall e'\neq e
     \right\} .
\label{eq:set2}
\end{equation}
Then, we evaluate
\begin{align} \label{eq:1edge}
  & L_e^*(A,\epsilon,P):=\max_{\tilde A\in \calC_e (A, \epsilon)} L(\tilde{A},P).
\end{align}
%

\begin{figure}[t]
     \centering
     \includegraphics[width=70mm]{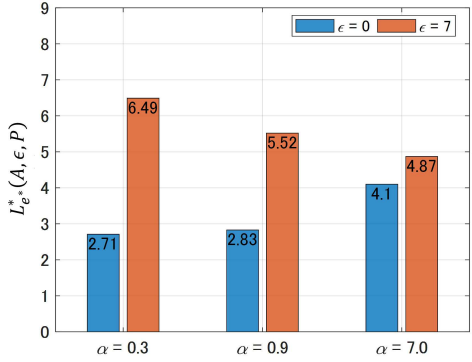}
     \caption{$L_{e^*}^*(A,\epsilon,P)$ with single-edge disruption of each $\alpha$ in $\epsilon=0$ and $\epsilon=7$.}
     \label{fig:bar_1edge}
\end{figure}

\begin{figure}[t]
     \centering
     \includegraphics[width=70mm]{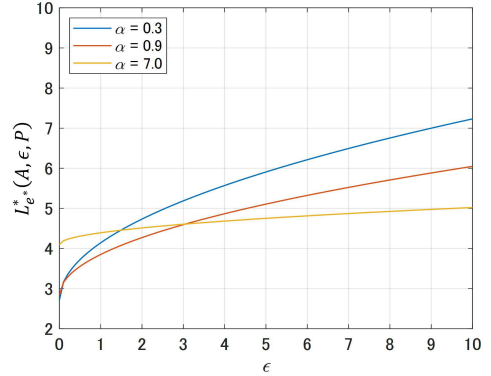}
     \caption{$L_{e^*}^*(A,\epsilon,P)$ with single-edge disruption of each $\alpha$ in each $\epsilon$.}
     \label{fig:plot_1edge}
\end{figure}

Here, we assume that some incident occurs at the edge $e^* := (f_1,w_4)$ in Fig.~\ref{fig:net} (e.g., large-scale traffic accidents, landslides, terrorism).
Fig.~\ref{fig:bar_1edge} shows $L_{e^*}^*(A,\epsilon,P)$ in (\ref{eq:1edge}) at each $\alpha$ for $\epsilon = 0$ and $\epsilon = 7$.
Furthermore, $L_{e^*}^*(A,\epsilon,P)$ of $\epsilon =7$ is reduced at all $\alpha$ values compared with Fig.~\ref{fig:total_cost} because the cost of only the single edge varies.
Notably, the difference in the worst costs between the design and after destruction is smaller than that in Fig.~\ref{fig:total_cost}.
As in Section \ref{sec:compa}, $L_{e^*}^*(A,\epsilon,P)$ with increasing $\epsilon$ is shown in Fig.~\ref{fig:plot_1edge}.
From Fig.~\ref{fig:plot_1edge}, we can observe that the fluctuation in $\alpha = 7.0$ is small.
Therefore, countermeasures for edge $(f_1,w_4)$ may be needed if the logistics manager implements an economic plan $\alpha=0.3$.
However, countermeasures will not be required if the resilience plan $\alpha = 7.0$ is selected.
Moreover, a high-risk road and base can be given by
\begin{align}\label{eq:w_edge}
  & \argmax_{e} L_e^*(A,\epsilon,P).
\end{align}


\section{Conclusion}
\label{sec:concl}

In this study, we formulated flexible logistics planning as an entropy-regularized optimization method and provided an efficient calculation method.
In addition, we proposed a resilience criterion based on probabilistic cost uncertainty, the scale of which is measured by KL divergence.
Through numerical simulations, we confirmed that entropy regularization can improve the resilience of the resulting logistic networks.

Although we only considered the 3-layer network, our results can be applied to any network. An evaluation of the proposed method based on real logistics network data is currently under investigation.
From a theoretical viewpoint, the direct optimization of the proposed resilience criterion is an interesting problem. In addition, because the fixed terminal density (e.g., $\zeta_s$) is restrictive in terms of representing realistic situations, we are interested in its relaxation via an \emph{unbalanced} optimal transport
theory; see, e.g., \cite{peyre2019computational}.


\bibliography{Resilient}

\begin{thebibliography}{18}
\providecommand{\natexlab}[1]{#1}
\providecommand{\url}[1]{\texttt{#1}}
\providecommand{\urlprefix}{URL }
\expandafter\ifx\csname urlstyle\endcsname\relax
  \providecommand{\doi}[1]{doi:\discretionary{}{}{}#1}\else
  \providecommand{\doi}{doi:\discretionary{}{}{}\begingroup
  \urlstyle{rm}\Url}\fi

\bibitem[{Betti et~al.(2022)Betti, Bezamat, Fendri, and Henkes}]{Forum2022}
Betti, F., Bezamat, F., Fendri, M., and Henkes, B. (2022).
\newblock Charting the course for global value chain resilience.
\newblock In \emph{World Economic Forum}.

\bibitem[{Chen et~al.(2016)Chen, Georgiou, Pavon, and
  Tannenbaum}]{chen2016robust}
Chen, Y., Georgiou, T., Pavon, M., and Tannenbaum, A. (2016).
\newblock Robust transport over networks.
\newblock \emph{IEEE transactions on automatic control}, 62(9), 4675--4682.

\bibitem[{Chen et~al.(2017)Chen, Georgiou, Pavon, and
  Tannenbaum}]{chen2017efficient}
Chen, Y., Georgiou, T.T., Pavon, M., and Tannenbaum, A. (2017).
\newblock Efficient robust routing for single commodity network flows.
\newblock \emph{IEEE Transactions on Automatic Control}, 63(7), 2287--2294.

\bibitem[{Darmawan et~al.(2021)Darmawan, Wong, and Thorstenson}]{darmawan2021}
Darmawan, A., Wong, H., and Thorstenson, A. (2021).
\newblock Supply chain network design with coordinated inventory control.
\newblock \emph{Transportation Research Part E: Logistics and Transportation
  Review}, 145, 102168.

\bibitem[{Dashkovskiy et~al.(2012{\natexlab{a}})Dashkovskiy, G{\"o}rges, and
  Naujok}]{Sergey2012}
Dashkovskiy, S., G{\"o}rges, M., and Naujok, L. (2012{\natexlab{a}}).
\newblock Autonomous control methods in logistics -- a mathematical
  perspective.
\newblock \emph{Applied Mathematical Modelling}, 36(7), 2947--2960.

\bibitem[{Dashkovskiy et~al.(2012{\natexlab{b}})Dashkovskiy, Karimi, and
  Kosmykov}]{Sergey2012_Raz}
Dashkovskiy, S., Karimi, H.R., and Kosmykov, M. (2012{\natexlab{b}}).
\newblock A {L}yapunov--{R}azumikhin approach for stability analysis of
  logistics networks with time-delays.
\newblock \emph{International Journal of Systems Science}, 43(5), 845--853.

\bibitem[{Davis et~al.(2021)Davis, Downs, and Gephart}]{davis2021towards}
Davis, K.F., Downs, S., and Gephart, J.A. (2021).
\newblock Towards food supply chain resilience to environmental shocks.
\newblock \emph{Nature Food}, 2(1), 54--65.

\bibitem[{Delvenne and Libert(2011)}]{delvenne2011}
Delvenne, J.C. and Libert, A.S. (2011).
\newblock Centrality measures and thermodynamic formalism for complex networks.
\newblock \emph{Physical Review E}, 83(4), 046117.

\bibitem[{Eysenbach and Levine(2021)}]{eysenbach2021}
Eysenbach, B. and Levine, S. (2021).
\newblock Maximum entropy rl (provably) solves some robust rl problems.
\newblock \emph{arXiv preprint arXiv:2103.06257}.

\bibitem[{Haarnoja et~al.(2018)Haarnoja, Zhou, Abbeel, and
  Levine}]{haarnoja2018soft}
Haarnoja, T., Zhou, A., Abbeel, P., and Levine, S. (2018).
\newblock Soft actor-critic: Off-policy maximum entropy deep reinforcement
  learning with a stochastic actor.
\newblock In \emph{International conference on machine learning}, 1861--1870.
  PMLR.

\bibitem[{Han et~al.(2020)Han, Chong, and Li}]{han2020}
Han, Y., Chong, W.K., and Li, D. (2020).
\newblock A systematic literature review of the capabilities and performance
  metrics of supply chain resilience.
\newblock \emph{International Journal of Production Research}, 58(15),
  4541--4566.

\bibitem[{Hatefi and Jolai(2014)}]{hatefi2014}
Hatefi, S.M. and Jolai, F. (2014).
\newblock Robust and reliable forward--reverse logistics network design under
  demand uncertainty and facility disruptions.
\newblock \emph{Applied mathematical modelling}, 38(9-10), 2630--2647.

\bibitem[{Ito and Kashima(2022)}]{kk2022}
Ito, K. and Kashima, K. (2022).
\newblock Sinkhorn {MPC}: Model predictive optimal transport over dynamical
  systems.
\newblock In \emph{2022 American Control Conference (ACC)}, 2057--2062.
\newblock \doi{10.23919/ACC53348.2022.9867406}.

\bibitem[{Ministry~of Economy(2021)}]{Trade2021}
Ministry~of Economy, T.a.I. (2021).
\newblock White paper on international economy and trade 2021.
\newblock \emph{White Paper on International Economy and Trade}.

\bibitem[{Mnih et~al.(2016)Mnih, Badia, Mirza, Graves, Lillicrap, Harley,
  Silver, and Kavukcuoglu}]{mnih2016}
Mnih, V., Badia, A.P., Mirza, M., Graves, A., Lillicrap, T., Harley, T.,
  Silver, D., and Kavukcuoglu, K. (2016).
\newblock Asynchronous methods for deep reinforcement learning.
\newblock In \emph{International conference on machine learning}, 1928--1937.
  PMLR.

\bibitem[{Nozick and Turnquist(1998)}]{nozick1998}
Nozick, L.K. and Turnquist, M.A. (1998).
\newblock Integrating inventory impacts into a fixed-charge model for locating
  distribution centers.
\newblock \emph{Transportation Research Part E: Logistics and Transportation
  Review}, 34(3), 173--186.

\bibitem[{Peyr{\'e} et~al.(2019)Peyr{\'e}, Cuturi
  et~al.}]{peyre2019computational}
Peyr{\'e}, G., Cuturi, M., et~al. (2019).
\newblock Computational optimal transport: With applications to data science.
\newblock \emph{Foundations and Trends{\textregistered} in Machine Learning},
  11(5-6), 355--607.

\bibitem[{Prorok et~al.(2021)Prorok, Malencia, Carlone, Sukhatme, Sadler, and
  Kumar}]{prorok2021}
Prorok, A., Malencia, M., Carlone, L., Sukhatme, G.S., Sadler, B.M., and Kumar,
  V. (2021).
\newblock Beyond robustness: A taxonomy of approaches towards resilient
  multi-robot systems.
\newblock \emph{arXiv preprint arXiv:2109.12343}.

\end{thebibliography}

\appendix
\section{Proof of Theorem \ref{the:sh}}    
\label{app:sh}

Let us relabel the elements of $\calV$ as $ \calV :=\{1,2,\cdots,n\}$ and define the matrix $\vec{B}\in \Rnn$ by
\begin{eqnarray*}
     \begin{split}
       \vec{B}_{(i,j)} := \left\{
         \begin{array}{l}
           \exp{\left(- \frac{\vec{A}_{(i,j)}}{\alpha}\right)} : \text{edge}(i,j) \in \mathcal{E} \\
           0 \hspace{21.5mm}: \text{edge}(i,j) \notin \mathcal{E}.
         \end{array}
        \right.
     \end{split}
\end{eqnarray*}
Let $\lambda_{B}$ be the maximum eigenvalues of $\vec{B}$, as well as $\vec{v} \ \text{and} \ \vec{u} $ be the corresponding right and left eigenvectors, respectively. Then, for $x=(x_0,x_1,\ldots,x_T)\in \calV^{T+1}$ (with $T=3$), we define:
\begin{eqnarray*}
     \begin{split}
          &\mathfrak{M}_{\rm RB} (x_0, \dots, x_{T}) := \\
          &\;\; {\vec{u}_{x_{0}}\vec{v}_{x_{T}}} \lambda_{B}^{- T} \exp{\left(- \frac{\sum_{t=0}^{T-1} \vec{A}_{(x_i,x_i+1)}}{\alpha}\right)},
     \end{split}
\end{eqnarray*}
where $\vec{v}_i$ denotes the $i$-th element of $\vec{v}$.
It can be verified that $\mathfrak{M}_{\rm RB}$ is a probability measure of $\calV^{T+1}$; see \cite{chen2016robust,chen2017efficient}.

For any probability measure $P$ on $\calV^{T+1}$ that satisfies
\begin{align}
  & P(x_0=\rm i) = 1, \label{eq:constApp1}\\
  & P(x_T=s) = \zeta_s,\ s\in \calS,\ P(x_T=v) = 0,\ v\notin \calS,
  \label{eq:constApp2}
\end{align}
we obtain
\begin{eqnarray*}
     \begin{split}
       & \KL{P}{\mathfrak{M}_{\rm RB}} \\
       &= \sum_{(x_0,\dots,x_{T})} P(x_0, \dots, x_{T}) \log{\left(\frac{P(x_0,\dots,x_{T})}{\mathfrak{M}_{\rm RB} (x_0,\dots,x_{T})}\right)} \\
       &= \frac{1}{\alpha} \left(\sum_{x_0,\dots,x_{T}} P (x_0,\dots,x_{T}) \sum_{t=0}^{T-1} \vec{A}_{(x_t,x_{t+1})} - \alpha \mathcal{H}(P)\right) \\
       & \quad - \log{\vec{u}_{\rm i}}
       - \sum_{s\in\calS} \zeta_s \log{\vec{v}_{s}}.
     \end{split}
\end{eqnarray*}
We note that only the first term depends on $P$ as long as \eqref{eq:constApp1} and \eqref{eq:constApp2} are satisfied.
This completes the proof of Theorem \ref{the:sh}.

\end{document}